\documentclass[twoside]{article}
\usepackage{fleqn,espcrc2}
\usepackage[dvips]{graphicx}
\usepackage[figuresright]{rotating}
\usepackage{epsf}
\usepackage{amssymb}
\def\bc{\begin{center}}
\def\ec{\end{center}}
\def\beq{\begin{equation}}
\def\eeq{\end{equation}}
\def\bs{\begin{slide}}
\def\es{\end{slide}}
\newcommand{\bmath}{\begin{displaymath}}
\newcommand{\emath}{\end{displaymath}}
\newcommand{\beqn}{\begin{eqnarray}}
\newcommand{\eeqn}{\end{eqnarray}}
\newcommand{\beqns}{\begin{eqnarray*}}
\newcommand{\eeqns}{\end{eqnarray*}}
\newcommand{\ba}{\begin{array}{c}} 
\newcommand{\bat}{\begin{array}{cc}} 
\newcommand{\ea}{\end{array}}

\newcommand{\lsim}{\stackrel{<}{_\sim}}

\newcommand{\gev}{\, \mbox{GeV}}

\title{Three-point Green Functions in the resonance region~:
LEC's
\thanks{Talk given at the High--Energy Physics International
Conference on Quantum Chromodynamics, 2-9 July (2005), Montpellier (France);
IFIC/05$-$40 report. To appear in the Proceedings.}
}

\author{J.~Portol\'es\address[Espanya]{Instituto de F\'{\i}sica Corpuscular,
IFIC, CSIC-Universitat de Val\`encia, \\ Apt. Correus 22085, E-46071 Val\`encia,
Spain}
} 
       
\begin{document}

\begin{abstract}
Using the $1/N_C$ expansion scheme and truncating the hadronic
spectrum to the lowest-lying multiplets of resonances we develop
a procedure to implement QCD information on the non-perturbative
regime populated by resonances through the use of three-point
Green Functions of QCD currents. The process implies to enforce
on these functions the constraints provided by their asymptotic behaviour
ruled by QCD using their OPE or associated form factors. As a result
we are able to determine several couplings of the ${\cal O}(p^6)$ 
Chiral Perturbation Theory Lagrangian and, in consequence, to address
phenomenological issues such as the $SU(3)$ breaking in $K_{\ell 3}$ 
decays.
\vspace{1pc}
\end{abstract}

\maketitle

\section{Introduction}

Quantum Chromodynamics (QCD) is, knowledgeably,
the theory of strong interactions.
However in spite of its early success in the description and 
understanding of the interaction between quarks and gluons, it soon
became clear that, due to its asymptotic freedom property, the study
of the low energy processes involving light-flavoured hadrons
(typically $E \lsim  2 \gev$) would not be possible with a 
strong interaction theory written in terms of quarks and gluons as 
dynamical degrees of freedom.
In the early eighties, and
relying in a very fruitful heritage from the pre-QCD era, 
it was observed 
that the chiral symmetry of massless QCD could be used to 
construct a strong interaction
field theory, intended to be dual to QCD, in terms of the 
lightest $SU(3)$ octet of pseudoscalar mesons in the role 
of pseudo-Goldstone bosons associated to the spontaneous 
breaking of that symmetry. This construction, known as 
Chiral Perturbation Theory ($\chi PT$) 
\cite{Weinberg:1978kz,Gasser:1984gg}, has been
very much useful in the study of strong interaction effects 
at very low energy, where the
theory has its domain, namely $E \ll M_V$ (being $M_V$ 
the mass of the $\rho (770)$,
the lightest hadron not included in the theory). 
Its success has pervaded
hadron physics in the last two decades, bringing to the main front 
the concept of effective field theory as a powerful tool to handle the 
non-perturbative regime of QCD. 
An effective field theory tries to embody the main features 
of the fundamental theory in order
to handle the latter in a specific energy regime where it is, 
whether more inconvenient or just
impossible, to apply it.
\par
The $\chi$PT Lagrangian has a perturbative structure guided by powers of the
soft external momenta and light quark masses. Hence one can construct
the theory up to a fixed order in the expansion, ${\cal O}(p^n)$, by 
adding to the previous terms the last order operators in
${\cal L}_n^{\chi PT}$, namely~:
\begin{equation}
{\cal L}_{\chi PT} \, = \, {\cal L}_2^{\chi PT} + {\cal L}_4^{\chi PT} +
{\cal L}_6^{\chi PT} + ...
\end{equation}
The first term ${\cal L}_2^{\chi PT}$ embodies the spontaneous breaking
of the chiral symmetry and depends only on two parameters~: $F$, the decay
constant of the pion, and 
$B_0 \, F^2 = - \langle 0 | \overline{\psi} \psi |0 \rangle $, the vacuum
expectation value of the light quarks; both of them in the chiral limit.
Higher orders in the expansion bring in the information from 
short-distance contributions that have been integrated out, for instance
resonance states. This information is incorporated into the low energy
couplings (LECs) that weight the operators of the theory~:
\begin{eqnarray}
{\cal L}_4^{\rm \chi PT} & = & \sum_{i=1}^{10} \, L_i \; {\cal O}_i^{(4)}
\; ,  \nonumber \\   
{\cal L}_6^{\rm \chi PT} & = & \sum_{i=1}^{90} \, C_i \; {\cal O}_i^{(6)}
\; , 
\end{eqnarray}
for three flavours. The explicit expressions for the operators can be
read in Refs.~\cite{Gasser:1984gg,Bijnens:1999sh}. The reference
scale in the chiral expansion $\Lambda_{\chi} \sim M_V$ indicates that
LECs in $\chi$PT should receive contributions from the energy regime
at or above this scale. In Ref.~\cite{Ecker:1988te} the contributions
of the lightest multiplets of resonances to the ${\cal O}(p^4)$ LECs in
${\cal L}_4^{\chi PT}$ were computed and it was observed that they 
saturated the values extracted from phenomenological analyses. Hence 
it is sound to think that the most important contribution to the LECs
is provided precisely by the energy region of the last integrated scale
($E \sim \Lambda_{\chi}$).

\section{The Resonance Chiral Theory}

The phenomenology of hadron physics at low energies is starting to 
involve more precise predictions and, consequently, a good knowledge on 
the LECs becomes necessary. At ${\cal O}(p^4)$
a thorough study was carried out in Refs.~\cite{Ecker:1988te,Ecker:1989yg}
using a phenomenological Lagrangian that is intended to carry the
information of QCD in the energy region of the hadronic resonances.
The guiding principles that are implemented in their construction are
the chiral symmetry, that describes the interactions of the Goldstone
bosons, and the flavour symmetry that is collected in the representations
of the resonance fields. This model-independent construction is called
Resonance Chiral Theory (R$\chi$T). Schematically the structure
of its operators is~:
\begin{equation}
{\cal O} \, \sim \,  \left\langle R ... R \, \chi(p^n) \right\rangle \, , 
\end{equation}
where $R$ is short for a unitary representation of the resonance fields
(the lightest multiplets of vector, axial-vector, scalar and pseudoscalar
resonances), 
$\chi(p^n)$ is a chiral tensor of ${\cal O}(p^n)$ in the $\chi$PT 
counting, and the trace operates in the flavour space. As happens in 
$\chi$PT the symmetries do not give any information on the couplings of
these operators that remain unknown from these settings. One of the
most relevant properties of this construction is that, upon decoupling
of the resonance fields, one recovers the structure of the $\chi$PT
Lagrangian, hence allowing the writing of the $\chi$PT LECs in terms of
the couplings of the R$\chi$T.
\par
At ${\cal O}(p^4)$ one can see \cite{Ecker:1988te} 
that the only operators contributing to the LECs are those with one only
resonance field and a chiral tensor ${\cal O}(p^2)$, 
$\langle R \chi(p^2) \rangle$. In the $U(3)$ limit only 6 operators 
with this structure appear. The construction of the R$\chi$T to 
obtain the ${\cal O}(p^6)$ chiral LECs is much more involved 
and it requires, together with the previous ones, 70 operators 
with the structure 
$\langle R \chi(p^4)  \rangle$, 38 operators 
$\langle R_1 R_2 \chi(p^2) \rangle$ and 7 operators with only resonance
fields $\langle R_1 R_2 R_3 \chi(p^0) \rangle$. These have been studied in 
Ref.~\cite{nos}. 
\par
We would like to emphasize that the use of R$\chi$T helps to establish 
relations between LECs and also parameterizes them in terms of the new
unknown couplings of the theory with resonance fields. However 
R$\chi$T, by itself, does not determine the LECs in $\chi$PT. Therefore more 
information will be needed. On the other side
R$\chi$T plays an important role in the implementation of the large-$N_C$
($N_C$ is short for the number of colours) ideas as we will see in the
following.

\section{LECs from Large-$N_C$ QCD}
\label{sect:3}

The determination of the $\chi$PT LECs asks for a thorough knowledge
of QCD in the energy region populated by light-flavoured resonances
($M_V \lsim E \lsim 2 \, \gev$). This is a highly non-trivial issue
that requires to master poorly known aspects such as bound and
resonant states, duality and hadronization mechanisms, i.e. 
non-perturbative QCD. It has been
known for long that the $SU(N_C)$ gauge theory in the 
$N_C \rightarrow \infty$ limit is
more simple than the real theory while still
resembling both qualitatively and quantitatively the $N_C=3$ case
\cite{'tHooft:1974hx}. 
The $1/N_C$ expansion tells us that, at leading order, any Green
Function of QCD currents (quark bilinears) can be given by a sum of all
possible topologies of tree level diagrams allowed by an infinite 
spectrum of zero-width mesons. Consequently the final result for a 
Green Function in the $N_C \rightarrow \infty$ limit is an
infinite sum of meromorphic functions (meson poles). Moreover
this structure is further constrained as it has to satisfy Ward identities
and other chiral symmetry restrictions for $E \ll M_V$ and
it also may match
the asymptotic behaviour ruled by perturbative QCD at $E \gg M_V$
\cite{Ecker:1989yg,Knecht:2001xc,Pich:2002xy}.
The interrelation between these different sources of information is 
the main goal of our project. Refs.~\cite{Knecht:1997ts} provide
some seminal work on the link between the Large-$N_C$ expansion
and resonance saturation.
\par
A general property of the LECs in $\chi$PT is that they are coefficients
of the Taylor expansion of some QCD Green Functions (once singularities
associated with the contributions of low-momentum pseudoscalar intermediate
states have been subtracted). The relevant correlation functions are,
precisely, order parameters of the spontaneous breaking of chiral
symmetry. Accordingly they do not receive contributions from perturbative
QCD in the chiral limit and one expects LECs to be sensitive to the 
physics of the intermediate energy region ($\Lambda_{\chi}$) 
\cite{Knecht:2001xc}. 
\par
Our proposal involves the use of the meromorphic approach to Green 
Functions, in terms of resonance or Goldstone states, in order to match 
its high-energy behaviour with the one provided
by the Operator Product Expansion (OPE) at leading order in $\alpha_S$
and the momenta series. 
Hence we get information that we can translate
into the LECs participating in the Green Function through its Taylor
expansion in momenta. Several technical details follow~:
\begin{itemize}
\item[-] In order to construct the meromorphic procedure we include the lightest
multiplets of resonances por $J^{PC}=0^{++}, 0^{-+}, 1^{--}, 1^{++}$
states, one for each channel. The cut in the spectrum means that
we are not implementing
properly the large-$N_C$ limit, though well known phenomenology suggests
that higher masses contributions tend to be much suppressed.
\item[-] The determination of the meromorphic approach to the Green Function
can be done in two different ways: on one side one can construct
an {\em ansatz} that has the right properties; on the other a phenomenological
Lagrangian, like R$\chi$T, can be employed to evaluate the 
Feynman diagrams that give
the different topologies at tree level. Though 
both procedures are germane the second has the advantage of extracting
information on the R$\chi$T too, that later can be used in other
settings.
\item[-] We try to match our construction of the Green Function with the
leading order in its OPE and, in addition we can also enforce the 
asymptotic behaviour of related form factors (their vanishing at high
transfer of momentum) as it results from the analysis of spectral functions
\cite{Ecker:1989yg}. It has been noticed \cite{Bijnens:2003rc} that, with a 
finite number of states, it may happen that inconsistencies between the
conditions from form factors and OPE appear.
When studying the  $\langle VAP \rangle$ and the $\langle S PP \rangle$  Green
functions ($V$ is short for vector, $A$ for axial-vector, 
$S$ for scalar and $P$ for pseudoscalar currents) we have concluded that
we were able to enforce both the asymptotic
behaviour, for form factors involving stable final states, and the OPE
constraints.
However the corresponding restriction on form factors with 
resonances could not be compelled. We will have to deal
with this feature that, in practice, turns out to be not too severe. 
\end{itemize} 
The application of this program in the study of three Green
functions of QCD currents, 
namely $\langle V V P \rangle$ \cite{Ruiz-Femenia:2003hm},
$\langle VAP \rangle$ \cite{Cirigliano:2004ue} and 
$\langle S PP \rangle$ \cite{Cirigliano:2005xn} has been very rewarding. 

\section{Example~: $SU(3)$ breaking in $K_{\ell 3}$ decays}

$K_{\ell 3}$ decays offer one of the most accurate determinations of 
the $V_{us}$ CKM element. The main uncertainty in extracting this 
matrix element comes from theoretical calculations of the 
vector form factor $f_{+}^{K^0 \pi^-}(0)$ defined by~:
\begin{eqnarray}
\langle \pi^- (p) | \bar{s} \gamma_\mu u | K^0 (q) \rangle & = &  
f_{+}^{K^0 \pi^-} (t)  \, (q + p)_\mu \nonumber \\
& &  \! \! \!  \! \! \! +   \, 
f_{-}^{K^0 \pi^-} (t)  \, (q - p)_\mu   \, , 
\end{eqnarray} 
with $t=(q-p)^2$. For $t=0$ deviations of $f_{+}^{K^0 \pi^-}(0)$ from
the unity (the octet symmetry limit) are of second order in $SU(3)$ 
breaking, hence
\begin{equation}
f_{+}^{K^0 \pi^-}(0) = 1 + f_{p^4} + f_{p^6} + ...
\end{equation}
The first correction is ${\cal O}(p^4)$ in $\chi$PT and it gives
$f_{p^4} = -0.0227$ with virtually no uncertainty 
\cite{Gasser:1984ux,Leutwyler:1984je}. At ${\cal O}(p^6)$ the situation 
is more complex as there appear contributions from two-loops with 
${\cal L}_2^{\chi PT}$, 
one-loop with one vertex from the ${\cal L}_4^{\chi PT}$ Lagrangian,
and tree-level
diagrams with ${\cal L}_4^{\chi PT}$ and ${\cal L}_6^{\chi PT}$.
These loops give \cite{Bijnens:2003uy}~:
\begin{eqnarray}
 f_{p^6}^{\rm loops} 
(M_\rho) 
&=&  0.0093 \pm 0.0005 \, .     
\end{eqnarray}
Next, the tree level piece reads~:
\begin{eqnarray}
f_{p^6}^{\rm tree} (M_\rho) & = & 
8 \frac{\left( M_K^2 - M_\pi^2 \right)^2}{F_\pi^2}  \, \times \\
& & \! \! \! \! \! \! \! \!  \! \! \! \!
\left[\frac{\left(L_5^r (M_\rho) \right)^2}{F_\pi^2} - 
C_{12}^r (M_\rho) - C_{34}^r (M_\rho) \right] , \nonumber
\end{eqnarray}
that we can see involves the ${\cal O}(p^4)$ $L_5$ and the 
${\cal O}(p^6)$ $C_{12}$ and $C_{34}$ LECs. In 
Ref.~\cite{Cirigliano:2005xn} we applied the procedure outlined
in Section~\ref{sect:3} for the study of the $\langle SPP \rangle$
Green function. We evaluated the contribution of one multiplet of 
scalar resonances of mass $M_S$ and one multiplet of pseudoscalar
resonances of mass $M_P$, obtaining~:
\begin{eqnarray} \label{eq:lcs}
L_5^{SP} & = & \frac{F^2}{4 \, M_S^2} \, , \; \; \; \; \; \; \; \;  
\; \; \,  
C_{12}^{SP} \, = \,  - \frac{F^2}{8 M_S^4}   \, , \\
C_{34}^{SP} & = &  \frac{3 \, F^2}{16 M_S^4} + \frac{F^2}{16} 
\left(\frac{1}{M_S^2} - \frac{1}{M_P^2} \right)^2 \; , \nonumber
\end{eqnarray}
that provide a strong cancellation between the different contributions
in $f_{p^6}^{\rm tree}$.
With the numerics discussed in that reference we get~:
\begin{eqnarray} 
f_{p^6}^{\rm tree} (M_\rho) &=& - 0.002  \pm 0.012 \; \nonumber  \\
f_{p^6} & = & 0.007 \pm 0.012 \, .   
\end{eqnarray}
The value of $f_{p^6}$ is in slight disagreement with that in 
Ref.~\cite{Leutwyler:1984je}.
We finally obtain~:
\begin{eqnarray}
f_{+}^{K^0 \pi^-} (0) &=&  0.984 \pm 0.012  \, \\
|V_{us}| & = &  0.2201 \pm 0.0027_{f_+(0)} \pm 0.0010_{\rm exp} \; . 
\nonumber 
\end{eqnarray}
The ${\cal O}(p^6)$ LECs in Eq.~(\ref{eq:lcs}) also appear
in the slope of the scalar form factor $\lambda_0$ \cite{Bijnens:2003uy} 
defined by 
$f_0(t) = f_+(0) ( 1+ \lambda_0 t /M_{\pi^+}^2 + ... )$ and 
we obtain~:
\begin{equation}
\lambda_0 = \left( 13 \pm 3 \right) \cdot 10^{-3} \; ,
\end{equation}
in excellent agreement with a recent experimental 
measurement \cite{Alexopoulos:2004sy}.

\vspace*{0.2cm} 
\noindent {\bf Acknowledgements}  \\
This work has been supported in part by
MCYT (Spain) under grant FPA2004-00996, by Generalitat Valenciana
(Grants GRUPOS03/013, GV04B-594 and GV05/015) and by HPRN-CT2002-00311 
(EURIDICE).



\begin{thebibliography}{99}


\bibitem{Weinberg:1978kz}
  S.~Weinberg,
  PhysicaA {\bf 96} (1979) 327;
  J.~Gasser and H.~Leutwyler,
  Annals Phys.\  {\bf 158} (1984) 142.
  
\bibitem{Gasser:1984gg}
  J.~Gasser and H.~Leutwyler,
  Nucl.\ Phys.\ B {\bf 250} (1985) 465.

\bibitem{Bijnens:1999sh}
J.~Bijnens, G.~Colangelo and G.~Ecker,
JHEP {\bf 9902} (1999) 020
[arXiv:hep-ph/9902437].

\bibitem{Ecker:1988te}
  G.~Ecker, J.~Gasser, A.~Pich and E.~de Rafael,
  Nucl.\ Phys.\ B {\bf 321} (1989) 311.

\bibitem{Ecker:1989yg}
  G.~Ecker, J.~Gasser, H.~Leutwyler, A.~Pich and E.~de Rafael,
  Phys.\ Lett.\ B {\bf 223} (1989) 425.
  
\bibitem{nos} V.~Cirigliano, G.~Ecker, M.~Eidemuller, A.~Pich, 
              R.~Kaiser and J.~Portol\'es, \lq \lq Towards a large-$N_C$
	      estimate of the ${\cal O}(p^6)$ chiral low energy couplings",
	      work in preparation.  
  
\bibitem{'tHooft:1974hx}
  G.~'t Hooft,
  Nucl.\ Phys.\ B {\bf 75} (1974) 461;
  E.~Witten,
  Nucl.\ Phys.\ B {\bf 160} (1979) 57.

\bibitem{Knecht:2001xc}
  M.~Knecht and A.~Nyffeler,
  Eur.\ Phys.\ J.\ C {\bf 21} (2001) 659
  [arXiv:hep-ph/0106034].
  
\bibitem{Pich:2002xy}
  A.~Pich, in Proceedings of the Phenomenology of Large $N_C$ QCD, 
  edited by R.~Lebed (World Scientific, Singapore, 2002), p.~239,
  arXiv:hep-ph/0205030.

\bibitem{Knecht:1997ts}
  M.~Knecht and E.~de Rafael,
  Phys.\ Lett.\ B {\bf 424} (1998) 335
  [arXiv:hep-ph/9712457];
  S.~Peris, M.~Perrottet and E.~de Rafael,
  JHEP {\bf 9805}, 011 (1998)
  [arXiv:hep-ph/9805442].

 

\bibitem{Bijnens:2003rc}
  J.~Bijnens, E.~Gamiz, E.~Lipartia and J.~Prades,
  JHEP {\bf 0304} (2003) 055
  [arXiv:hep-ph/0304222].
  
\bibitem{Ruiz-Femenia:2003hm}
  P.~D.~Ruiz-Femenia, A.~Pich and J.~Portol\'es,
  JHEP {\bf 0307} (2003) 003
  [arXiv:hep-ph/0306157].
     
\bibitem{Cirigliano:2004ue}
  V.~Cirigliano, G.~Ecker, M.~Eidemuller, A.~Pich and J.~Portol\'es,
  Phys.\ Lett.\ B {\bf 596} (2004) 96
  [arXiv:hep-ph/0404004].

\bibitem{Cirigliano:2005xn}
  V.~Cirigliano, G.~Ecker, M.~Eidemuller, R.~Kaiser, A.~Pich and J.~Portol\'es,
  JHEP {\bf 0504} (2005) 006
  [arXiv:hep-ph/0503108].

\bibitem{Gasser:1984ux}
  J.~Gasser and H.~Leutwyler,
  Nucl.\ Phys.\ B {\bf 250} (1985) 517.
  
\bibitem{Leutwyler:1984je}
  H.~Leutwyler and M.~Roos,
  Z.\ Phys.\ C {\bf 25}, 91 (1984).

\bibitem{Bijnens:2003uy}
  J.~Bijnens and P.~Talavera,
  Nucl.\ Phys.\ B {\bf 669}, 341 (2003)
  [arXiv:hep-ph/0303103].
  
\bibitem{Alexopoulos:2004sy}
  T.~Alexopoulos {\it et al.}  [KTeV Collaboration],
  Phys.\ Rev.\ D {\bf 70} (2004) 092007
  [arXiv:hep-ex/0406003].



\end{thebibliography}
\end{document}